%% file: main.tex
\documentclass[sigconf,10pt,letterpaper]{acmart}
\usepackage{graphicx}
\usepackage{enumitem} 
\usepackage{multirow}
\usepackage{multicol}
\usepackage{xspace}

\newcommand{\system}{\textit{RARR} \xspace}

\begin{document}

\title[Robust Real-World Activity Recognition]{\system: Robust Real-World Activity Recognition with Vibration by Scavenging Near-Surface Audio Online}

\author{Dong Yoon Lee}
\affiliation{
    \institution{University of California, Merced}
    \city{Merced}
    \country{United States}}

\author{Alyssa Weakley}
\affiliation{
    \institution{University of California, Davis School Of Medicine}
    \city{Sacramento}
    \country{United States}}

\author{Hui Wei}
\affiliation{
    \institution{University of California, Merced}
    \city{Merced}
    \country{United States}}

\author{Blake Brown}
\affiliation{
    \institution{University of California, Davis School Of Medicine}
    \city{Sacramento}
    \country{United States}}

    \author{Keyana Carrion}
\affiliation{
    \institution{University of California, Davis School Of Medicine}
    \city{Sacramento}
    \country{United States}}

\author{Shijia Pan}
\affiliation{
    \institution{University of California, Merced}
    \city{Merced}
    \country{United States}}

\renewcommand{\shortauthors}{Lee et al.}

\begin{abstract}
One in four people dementia live alone, leading family members to take on caregiving roles from a distance. Many researchers have developed remote monitoring solutions to lessen caregiving needs; however, limitations remain including privacy preserving solutions, activity recognition, and model generalizability to new users and environments. Structural vibration sensor systems are unobtrusive solutions that have been proven to accurately monitor human information, such as identification and activity recognition, in controlled settings by sensing surface vibrations generated by activities. However, when deploying in an end user's home, current solutions require a substantial amount of labeled data for accurate activity recognition. Our scalable solution adapts synthesized data from near-surface acoustic audio to pretrain a model and allows fine tuning with very limited data in order to create a robust framework for daily routine tracking.
\end{abstract}

\maketitle

\input{sections/intro}
\input{sections/system_overview}
\input{sections/evaluation}
\input{sections/related_work}

\input{sections/discussion}
\input{sections/conclusion}

\bibliographystyle{ACM-Reference-Format}
\bibliography{acmart,shijia}

\end{document}

%% file: sections/intro.tex
\section{Introduction}
Many older adults desire to live independently as they age, including those with cognitive impairment. To help people with cognitive impairment age at home, adult children often take on caregiving roles from a distance. However, a major limitation to remote caregiving is the lack of awareness of health-related activities, such as taking medication, mobility, and sleeping, resulting in constant caregiver worry and crisis-driven care due to the lack of timely and informed intervention.



To automate monitoring of home activities, researchers have developed different sensor systems suitable for private indoor use, such as infrared/light-, RF-, and structural vibration-based sensors \cite{Yin2021May, nguyen2018eyelight, Liang2023Jun, zhang2023survey, hu2020fine, hu2022vma}. 
However, these privacy-preserving sensor systems face unique challenges compared to vision- or audio-based modalities, as they must indirectly infer information from inherently noisy data. 
Since their signals are susceptible to variations introduced by multiple physical factors, such indirect sensing systems often require large amounts of labeled data to achieve robust recognition.
However, such labeled data are typically scarce and annotation is particularly difficult because indirect sensing produces signals that are not interpretable by humans, unlike vision- and audio-based data.
Recent developments on foundation models \cite{Baris2025Jan, Wei2025Jun} 
introduce a new perspective for general representation models using self-supervised learning, where micro foundation models 
are developed for IoT sensing systems, such as vibration-based systems \cite{kimura2024efficiency, kimura2024vibrofm}. 


\begin{figure}[!t]
    \centering
    \includegraphics[width=0.8\linewidth]{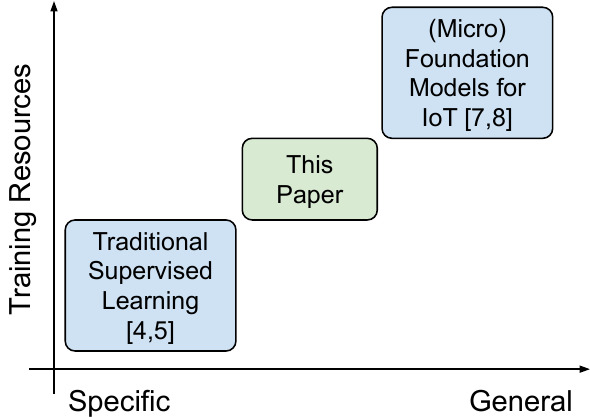}
    \caption{Scope of this work.}
    \vspace{-3ex}
    \label{fig:scope}
\end{figure}


Our Robust Real-World Activity Recognition (\system) solution explores real-world human activity recognition (HAR) using structural vibration data. 
\system falls within the scope between supervised learning and pre-trained foundation models in terms of training resources and specificity, as shown in Figure \ref{fig:scope}.
In order to keep the generalizations found in foundational models,
while optimizing for the activity recognition task like traditional supervised learning solutions,
we must use a model that can learn for both.
Variational autoencoders (VAEs) \cite{Zhao2021May} achieve robustness by separating the variational data from stochastic samples as variational latent, and achieves activity recognition task with mean latent. 
However, training VAEs with a small annotated vibration dataset will not accurately map the real world variational effects, such as from different users and environments, leading to lower accuracy.
For VAEs to accurately map the real world variation, we carefully curate an annotated dataset from audio data that is more widely available and pretrain a VAE with it.
To remove the audio modality bias, we finetune the pretrained VAE with our smaller vibration dataset.
However, it is challenging to preserve the robustness of the pretrained model when finetuning. 
To preserve this robustness, we use multitask learning to learn each latent separately, allowing the VAE to finetune only the mean latent with the activity recognition task without affecting the variational latent.

To effectively pretrain the VAE for in-home activity recognition task, \system curates a \textbf{near-surface} audio dataset of the target home activities.
We extract near-surface audio by searching for publicly available Autonomous Sensory Meridian Response (ASMR) recordings from different creators on the Internet.
Then we fine-tune the model with a small amount of labeled \textbf{on-surface} vibration data from a real world deployment.
To verify the performance of this solution, we compare the activity recognition accuracy of the finetuned model against three baselines to achieve a $13\%$ average improvement.

The contribution of this work includes:

\setlist{nolistsep}
\begin{itemize}[noitemsep]
\item Introducing a new framework for pretraining a model by curating modality- and task-aligned dataset from publicly available audio data.
\item Designing a multitask VAE that learns an activity recognition model robust to real-world data variance.
\item Developing a task-selective fine-tuning strategy for effective modality transfer.
\item Evaluating the system with real-world data collection and outperforming multiple baselines.
\end{itemize}



%% file: sections/system_overview.tex
\section{System Overview}

\begin{figure}
    \centering
    \includegraphics[width=.95\linewidth]{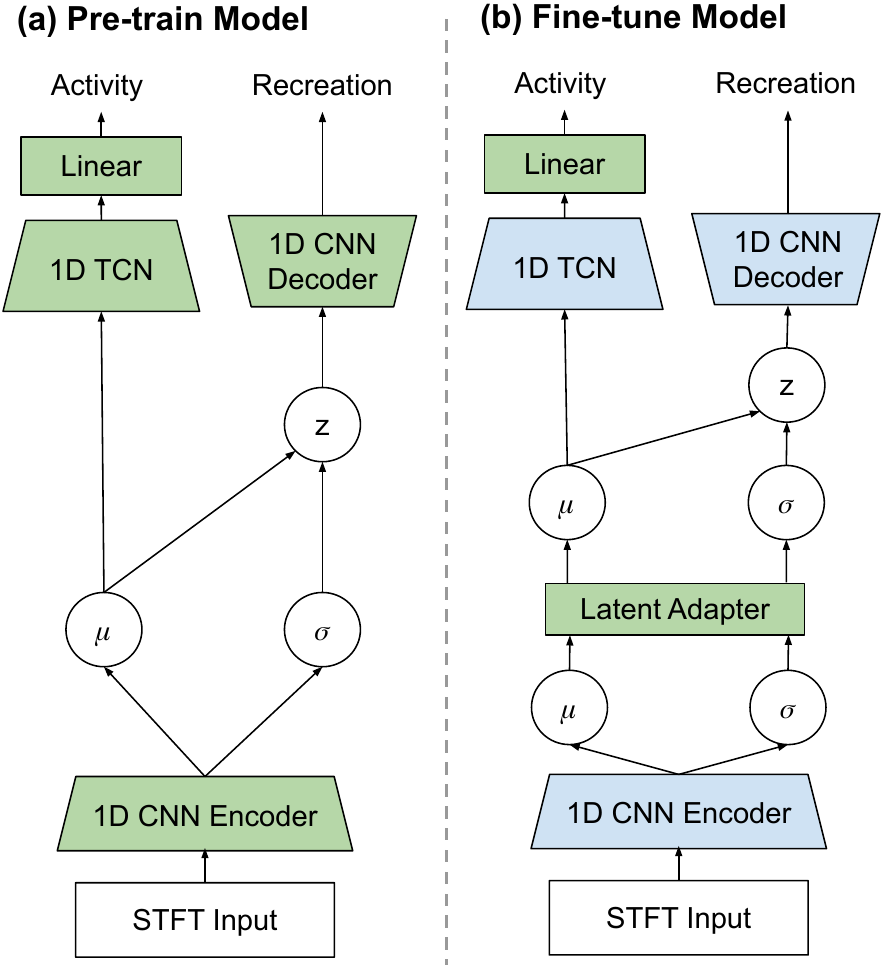}
    \caption{\system models.
    (a) The system first pre-trains a multitask VAE model on a large-scale ASMR audio dataset.
    (b) To adapt to the vibration modality, we introduce a low-rank 1D CNN adapter layer that fine-tunes the modality-specific latent parameters $\mu$ and $\sigma$. 
    In addition, we fine-tune the final linear layer of the temporal convolutional network (TCN) to account for modality-related distribution shifts. 
    All other modules remain frozen to retain the robustness learned from the ASMR pretraining. 
    }
    \label{fig:system-overview}
\end{figure}

To enable robust vibration-based activity recognition for different individuals with a small amount of labeled data, \system leverages near-surface ASMR audio data available on the Internet to pre-train an acoustic model of the target activities.
However, there is a modality gap between near-surface acoustic signals and on-surface acoustic signals, which makes the model not efficient for on-surface vibration signal prediction directly.
To tackle this challenge, we introduce the ASMR-Vibration Adapter (AVA), as a mechanism to account for the modality gap.
This allows us to fine-tune the pre-trained model with only a small amount of on-surface human-induced vibration data to retain the robust activity recognition capability.
Figure \ref{fig:system-overview} depicts \system system overview. 
The pre-trained model takes the short time Fourier transform (STFT) of the larger ASMR audio as the input and we form it as a multitask learning problem with two tasks of activity prediction and STFT reconstruction.
We adopt the VAE due to its ability to capture the essential target characteristics while remaining robust to variations.
Then the smaller structural vibration-based dataset is used to fine-tune the pre-trained model to mitigate the modality bias between the near-surface ASMR audio and the on-surface vibration signals.


\begin{figure}[!t]
    \centering
    \includegraphics[width=0.99\linewidth]{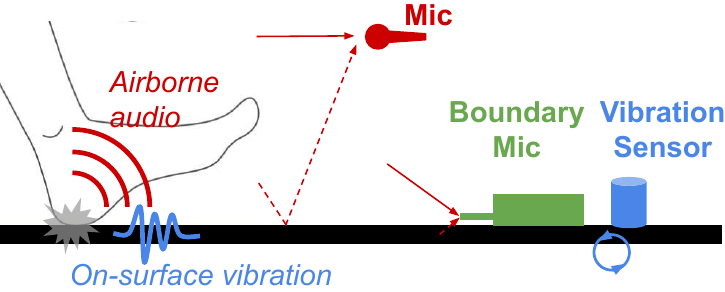}
    \caption{Near-surface audio and on-surface vibration.}
    \vspace{-3ex}
    \label{fig:ASMR}
\end{figure}

\subsection{Near-Surface ASMR Audio Dataset for On-Surface Vibration-based HAR}
Compared to common sensing modalities, such as computer vision and speech, human-induced vibration datasets are very limited.
With limited training data, it is difficult to establish HAR models that are robust to human and environment variance.
Therefore, we look into other forms of acoustic data available on the Internet.
Figure \ref{fig:ASMR} depicts the acoustic signals generated when a footfall occurs on the floor -- airborne audio and on-surface vibration.
When the foot strikes the floor, it induces vibrations dominated by the Rayleigh wave \cite{pan2017surfacevibe}, as marked in blue icons.
When microphone is used for room sound collection, as depicted with red icons, the reflected sounds often cause the comb filter effect and therefore impacts the sound quality.

High quality of room sounds on rigid surfaces are often collected by boundary microphones (a.k.a pressure zone mic), where the microphone diaphragm is placed very close (a few millimeters) to the surface, as shown in green icons in the figure.
When the sound wave reaches the rigid surface, it will create a ``pressure zone" between the diaphragm and the surface, where the direct and reflected waves are coherently in phase and reinforce each other. 
ASMR audio -- captured by boundary microphones with enhanced clarity for surface sounds \cite{asmr} -- are available on the Internet, including ASMR sounds of HAR at home.
To effectively generate the ASMR-HAR label dataset, we use keyword queries on popular search engines to link a set of activity labels to ASMR videos and audio.
We use this generated dataset to pretrain the encoders and decoders for both the activity recognition temporal convolutional network (TCN) and signal recreation decoder.

\subsection{Multitask Variational Autoencoder for Robust Activity Recognition}
In order to separate probabilistic information, such as user or environmental variance induced data distribution shift, from the categorical activity information, we adopt the VAE \cite{Doersch2016Jun}.
Instead of encoding the input $x$ into a single latent vector $z$, the encoder produces two outputs $\mu$ and $\sigma$, defining a Gaussian distribution $z \sim  \mathcal{N}(\mu, \sigma)$ \cite{Doersch2016Jun}.

In order to train only the modality bias shift when transfer learning between different modalities, it is essential to preserve the variance latent mapping of the large, pretrained data to retain robustness to different users and environmental changes. Therefore, we target this learning to limit any variance latent change by separating two learning tasks that each target different latents: 1. activity recognition with only the mean latents to isolate the modality specific information and 2. recreation to target the variance latents, so that the variance latents represent signal differences created by variables other than modality, such as user behavior and environment.
In this way, we can target the finetuning by unfreezing layers associated with the activity recognition task to shift the modality bias toward on surface data, while adding a adapter layer to account for slight differences in separation between near and on surface sensing.

To achieve the two tasks for separable latent information, we use a CNN encoder that takes the input STFT signal to generate the separated mean latent ($\mu$) and variance latent ($\sigma$) information.
For the first task of activity recognition, a temporal convolutional network (TCN) takes only the mean latent ($\mu$) to find modality specific patterns across time and predicts an activity label.
For the second task of STFT signal recreation, the data recreation decoder takes the sample $z$ drawn from this distribution $\mu+\sigma \cdot \epsilon$, where $\epsilon$ is a random noise vector sampled from a standard normal distribution, to recreate the signal.
This design enables the separation of variation information, such as environmental or user variance that would effect the signal data, from the features that identifies the activities.


\subsection{Task-Selective Fine-Tuning for Efficient Modality Transfer}


In order to transfer the ASMR audio-based model for vibration-based activity recognition, we fine-tune the pretrained model with a small amount of vibration signal.
The multitask VAE design allows us to conduct the model transfer in an efficient way by only fine-tuning the supervised learning task.
This is because the supervised learning task relies on the mean latent state as input, and fine-tuning only this task enables the pretrained model to adapt to modality shifts on the supervised task, while preserving robustness to broader variance (e.g., across people, environments, and devices) captured from the larger dataset. 

As shown in Figure \ref{fig:system-overview}, we freeze (1) the shared encoder that extracts the VAE latent states from raw signals, (2) recreation decoder, and (3) TCN encoder for activity recognition, and fine-tune the last few layers for the activity recognition task only.
In order to account for changes of representation of modality related data used for the activity recognition task, as well as variational data that is used for the reconstruction, we add a low rank latent adapter during finetuning, specifically a multilayer 1D CNN, to shift the representation profile from the near-surface modality to the on-surface modality.

%% file: sections/evaluation.tex
\section{Evaluation}
We implement the end-to-end system and evaluate \system with data collected from a real-world deployment. 
We conduct experiments to demonstrate the advantage of our design in handling data variance and modality bias between pretrain and finetuning models.
Activity prediction accuracy
is used as the metric to evaluate the baselines and our approach.
We adopt three baseline models, including 
(1) \textit{Simple-VAE}: traditional supervised learning model with multitask VAE, 
(2) \textit{Pretrained-VAE}: multitask VAE model pretrained with ASMR audio data and without finetuning, 
and (3) \textit{A2V-VAE}: Pretrained-VAE finetuned with vibration data. 
Our \system can be considered as Pretrained-VAE task-selective finetuned with vibration data. 
The comparison of these baselines will allow us to understand the contribution of multitask VAE and task-selective finetuning respectively.




\subsection{Datasets}
We generate the ASMR dataset for training and collect human activity induced vibration in realistic settings for fine-tuning and testing.
We consider three common and important home activities for aging-in-place monitoring: walking, showering, medication intake, and medication refilling.

\paragraph{ASMR Dataset}
Publicly available ASMR audio data is collected.  
In order to align the context between the ASMR data with the aging-in-place application, we search for relevant keywords on popular search engines to obtain the data. 
For example, searching ``ASMR walking no talking'' will obtain acoustic files online labeled as ``walking'' in the application. 
By matching the application activity labels with the search terms, we compile a complete dataset from different creators and environments, which will enable robust training to recognize activities.
The ASMR dataset has a total of 3066 samples of 30 second audio that overlap by 15 seconds. These audio clips were aggregated from at least 3 ASMR videos per activity label curated by various search engines.
Due to the various availability of data for different activities, we balance the dataset such that each activity label has an equal amount of samples.

\paragraph{Home Health Simulation Suite Dataset}
We collected realistic home activity induced vibration signal in the University of California Davis' Home Health Simulation Suite.
The Simulation Suite is a one-bedroom apartment test-bed laid out in a U-shape and includes a living room, bedroom, bathroom, and kitchen. 
There are video cameras and microphones throughout the Simulation Suite which allows us to reliably establish ground truth by matching activity labels with vibration signals. 
We set up one geophone sensor on the floor in each room and an extra sensor on the bathroom counter to capture medication taking behavior.
Four participants were instructed to conduct 30 different sequences involving 4 (of possible 23; including walking, showering, medication taking, and medication refilling) activities.


\subsection{Experiment Design}
To pretrain the \textit{Pretrained-VAE} model, we split the ASMR dataset to $60\%$ training and $40\%$ validation.
We finetune the pretrained model with the vibration data of all five activity sequences from participant 1 to generate the \system model and \textit{A2V-VAE} model.
We also train a supervised learning based VAE -- \textit{Simple-VAE} -- from scratch using the same vibration data from participant 1 without pretraining.
To demonstrate the effectiveness of our designs, we test the model using vibration data from other participants 2, 3, and 4 to show the robustness of the model on unseen users with only very small amount of labeled data from one person.

\begin{figure}
    \centering
    \includegraphics[width=0.90\linewidth]{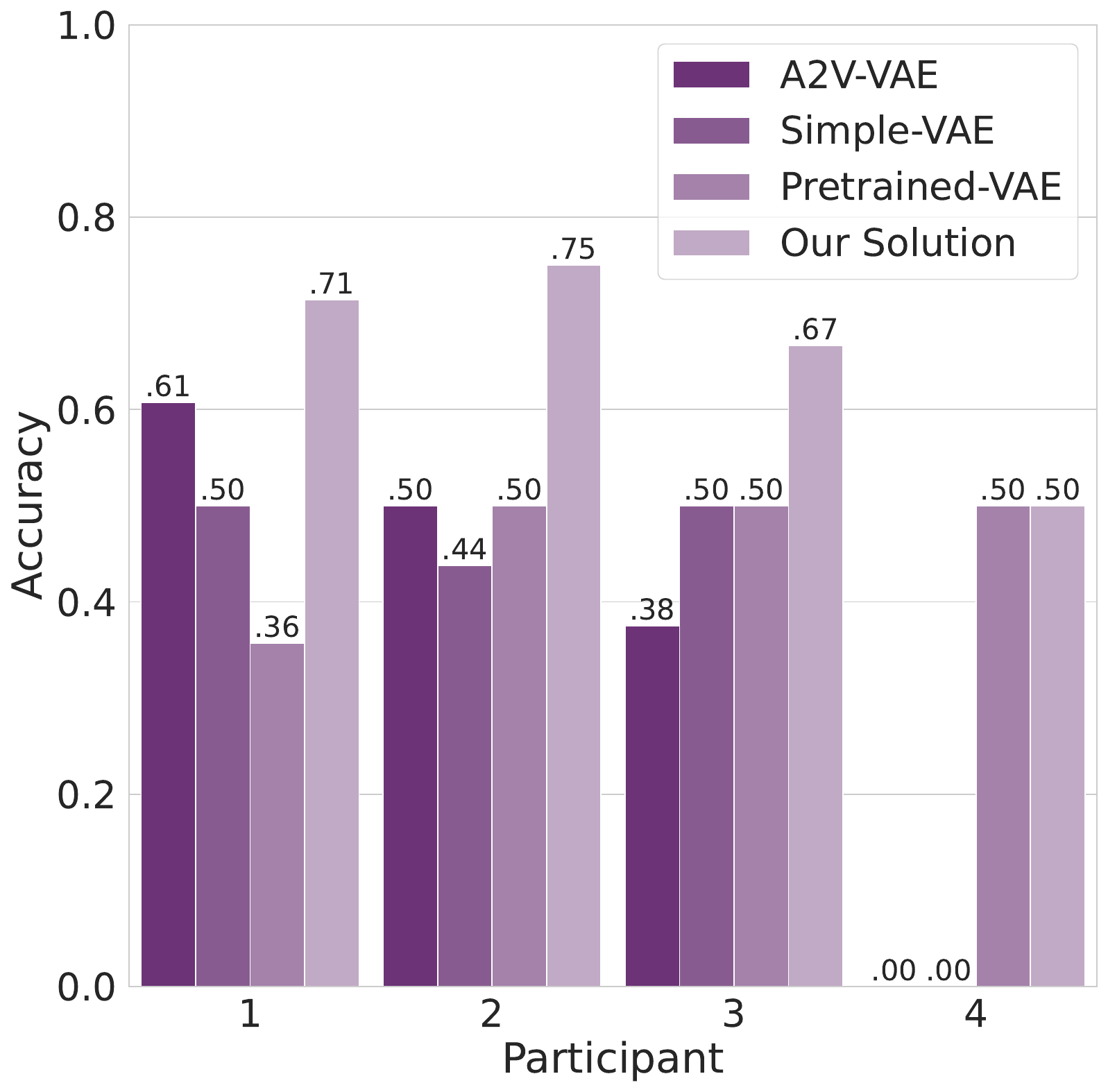}
    \caption{Activity recognition accuracy of \system and three baselines.
    }
    \vspace{-3ex}
    \label{fig:results:person}
\end{figure}

\subsection{Results Analysis}


Figure \ref{fig:results:person} depicts the activity recognition accuracy. 
We show 4 different models, with \textit{A2V-VAE} in dark purple bars, \textit{Simple-VAE} in medium dark purple bars, and \textit{Pretrained-VAE} in light purple bars as our baselines, in comparison with our solution, \system in the lightest purple bar.
We group the accuracy metrics by participant in the x-axis and show the accuracies in the y-axis.
Since we finetune the \textit{A2V-VAE} and \textit{RARR} (our solution) and trained the \textit{Simple-VAE} with participant 1's data, we consider the first column as the training accuracy and the rest as the testing accuracy for these models.
Unlike traditional supervised learning models that overfit the activity recognition task, these three multitask solutions balance the loss of both tasks (the activity recognition task and the recreation task), such that the recognition task does not overfit to help with the recreation task. 
We consider all columns are testing accuracy for \textit{Pretrained-VAE}.

For the overall accuracy, our \system achieves the highest accuracy for all participants ($71\%$, $75\%$, $67\%$, $50\%$), with an overall mean accuracy of $72\%$ and the standard error of $18\%$. 
The best performing baseline, \textit{A2V-VAE}, performs worse for all participants ($61\%$, $50\%$, $38\%$, $0\%$), possibly because when it finetunes with the vibration dataset, it does not retain the robustness it had learned when pretraining on the ASMR dataset.
\textit{Simple-VAE} performs worse than \textit{A2V-VAE} on average ($50\%$, $44\%$, $50\%$, $0\%$).
Not pretraining on the ASMR dataset constrains \textit{Simple-VAE} to only learn the limited user variance from the small vibration dataset of one person.
Although \textit{Pretrained-VAE} shows the worst performance on average compared to all other models ($36\%$, $50\%$, $50\%$, $0\%$), we see a higher consistency between participants compared to other baselines, especially Participant 4.
This indicates that the pretraining step with ASMR dataset is essential to model user variance.

%% file: sections/related_work.tex
\section{Related Work} 

\subsection{Acoustic-based Human Activity Recognition (HAR)}
The acoustic signals used for HAR mainly include airborne audible sounds \cite{laput2018ubicoustics, liang2019audio}, ultrasounds, and structural vibration signals.
Audio-based solutions leverages audible sounds generated through everyday activities for HAR.
Professional sound effect libraries in entertainment industry \cite{laput2018ubicoustics} and online videos \cite{liang2019audio} have been adopted for activity of daily living (ADL) recognition.
However, audible sounds often contain sensitive information that poses privacy challenges when used for continuous monitoring at home.
To mitigate the privacy concerns, ultrasound sensors were adopted to capture human motion via active sensing \cite{murray2017bio, ghosh2023ultrasense, tanigawa2024hear}.
However, ultrasound-based approaches require active signal emission, which not only increases power consumption but may also disturb household animals that are sensitive to ultrasonic frequencies.
Structural vibration signals, as a privacy-preserving passive sensing alternative, were explored for occupant monitoring \cite{hu2020fine, hu2022vma}, and we focus on the robustness challenge posed by limited labeled data in this work.

\subsection{Acoustic-based Foundation Models}
With the development of foundation models, vibration-based micro foundation models have been explored \cite{kimura2024efficiency, kimura2024vibrofm}. 
Self-supervised techniques are introduced to pretrain a foundation model that outputs effective representation using a large amount of heterogeneous acoustic signals and can be finetuned for downstream tasks.
The encoder can be moderately sized and real-time inference on edge is also explored with Raspberry Pi devices \cite{kimura2024case}.
This work is inspired by the acoustic-based micro foundation models and explores pretrained acoustic representation models from a modality-proximity perspective to enable robust performance with relatively smaller models and computational needs.

%% file: sections/discussion.tex
\section{Discussion}

\subsection{Making Autoencoders More Robust}
Autoencoders learn to map the distribution for a given feature space in order to stochastically represent different samples, such as images or signals.
However, there are many other improvements to autoencoders that can further enhance feature encoding and robustness for real world applications.
For example, masked autoencoders use various prepossessing masks to limit the data needed to extract useful features and therefore increase performance.
Other problems with VAE may include compressing too much activity and modality information into the latents such that you may lose important information.
Conditional autoencoders can solve for this compression problem by using conditional latents, such as class labels, to further reduce the feature space by shifting the distribution according to these conditions.



\subsection{Long Term In-Home Deployment Challenges}
While we demonstrate the proof-of-concept that \system can be used to recognize real-world activity both within and between participants, the true utility and robustness of this system will be tested when we perform larger scale community deployments in a range of built environments. For any smart system to be scalable, it must be generalizable with a high degree of accuracy, sensitivity, and specificity; this is particularly true for an activity recognition system that aims to reduce caregiver worry and crisis driven care. Thus, it will be important that false negatives (not detecting a health event) are balanced against false positives (erroneously alerting caregivers). Additional challenges include managing normal within person variability versus increased variability that occurs in individuals with cognitive impairment and how to utilize this system both as a way to track cognitive health and intervene through just-in-time prompting to support independence. 


%% file: sections/conclusion.tex
\section{Conclusion}
This work introduces \system, a framework for robust real-world activity recognition using structural vibration. 
The system leverages online near-surface ASMR audio to pretrain models resilient to real-world variance, and then fine-tunes them on privacy-preserving on-surface vibration data.
By combining multitask variational autoencoding with a modality adaptation strategy, \system is able to effectively acquire acoustic representation for home activity prediction. 
Experimental evaluation demonstrates around $13\%$ improvement over baseline approaches, highlighting the potential of modality-adaptive transfer learning for privacy-preserving sensing systems. 
These results suggest that \system provides a practical path toward scalable deployment of unobtrusive activity recognition in everyday living environments, supporting independent aging and reducing caregiver worry.